\documentclass[aps,pra,twocolumn,groupedaddress]{revtex4} 

\usepackage{graphicx}
\usepackage{bm}
\usepackage{array}
\usepackage{amsmath}
\usepackage{amssymb}
\usepackage{subfigure}
\usepackage{ulem}
\usepackage{float}

\begin{document}


\title{Rabi Oscillations between Atomic and Molecular Condensates Driven with Coherent One-Color Photoassociation}



\author{ Mi Yan$^*$, B. J. DeSalvo$^*$, Ying Huang$^*$, P. Naidon$^\dag$, and T. C. Killian$^*$,}

\affiliation{$^*$Rice University, Department of Physics and Astronomy, Houston, Texas, USA 77251  \\
$^\dag$Quantum Hadron Physics Laboratory, Nishina Centre, RIKEN, 2-1 Hirosawa, Wako, Japan 351-0198}



\date{\today}

\begin{abstract}
We demonstrate coherent one-color photoassociation of a Bose-Einstein condensate, which results in
Rabi oscillations between atomic and molecular condensates. We attain atom-molecule Rabi frequencies that are comparable to decoherence rates by driving photoassociation of atoms in an $^{88}$Sr condensate to a weakly bound level of the metastable $^{1}S_{0}$+$^{3}P_{1}$ molecular potential, which has a long lifetime and a large Franck-Condon overlap integral with the ground scattering state.  Transient shifts and broadenings of the excitation spectrum are clearly seen at short times, and they create an asymmetric excitation profile that only displays Rabi oscillations for blue detuning from resonance.

%

\end{abstract}


\maketitle




Coherent conversion of atoms into molecules in quantum degenerate gases provides a path to create  molecular condensates, and  it is also of fundamental interest because of
 the complex dynamics of  many-body systems with nonlinear couplings \cite{dst04}. It has been demonstrated with  magnetic-field ramps near a magnetic Feshbach resonance (MFR) \cite{cgj10,kgj06} and with different configurations of two-color photoassociation (PA) \cite{wtt05,rbm04,spg12,jtl06}. Coherent one-color PA has typically been considered inaccessible due to the short lifetime of excited molecular states in alkali-metal atoms \cite{ntj08}. In addition, the observation of giant Rabi oscillations between atomic and molecular condensates has remained elusive, although it has been discussed extensively since the earliest days of the study of quantum gases \cite{tth99,ttc99,jma99,hwd00,icn04,kmc00,ggr01,jma02,gas04,gas04rapid,nse06,ntj08}.  Here, we demonstrate coherent, one-color PA and
Rabi oscillations between atomic and molecular condensates. We also observe transient  shifts and broadenings of the excitation spectra at short  times and indications of  universal dynamics on resonance \cite{ntj08}, all of which  have not been experimentally studied previously.
We access the coherent regime  by photoassociating atoms in an $^{88}$Sr condensate \cite{mmy10} to a weakly bound level of the metastable $^{1}S_{0}$+$^{3}P_{1}$ molecular potential. This
yields a long molecular lifetime and a large Franck-Condon overlap integral between ground and excited states, which allows atom-molecule coupling to exceed loss and decoherence rates.

The first study of coherent atom-molecule conversion \cite{dct02} used a Ramsey-pulse sequence with a MFR in a Bose-Einstein condensate (BEC) to create superpositions of atomic and molecular fields and  detect  atom-molecule oscillations. 
Rabi oscillations have been seen with MFRs in thermal bosons \cite{thw05}, bosonic-atom pairs  in a Mott insulator state in an optical lattice \cite{sbl07}, and a Bose-Fermi mixture \cite{opc09}.
Magnetoassociation is now a common tool for  creating  quantum degenerate molecular gases  and exploring the  unique properties of  universal Feshbach molecules \cite{kgj06,kgj03,thw05,ktj05}.

PA \cite{jtl06} is another well-established technique for forming ultracold molecules, and it is  fundamentally analogous to magnetoassociation \cite{cgj10}. Two-color PA, in which two light fields couple atomic scattering states to a ground molecular state through an intermediate, excited molecular level, is often used to create molecules in quantum degenerate gases through stimulated Raman transitions \cite{wfh00,rbm04}, creation of coherent atom-molecule superpositions  in a dark state \cite{wtt05}, or stimulated Raman adiabatic passage (STIRAP) \cite{bts98,spg12}.
These two-color processes are typically coherent, as demonstrated, for example, by the suppression of PA loss \cite{wtt05}.  Rabi oscillations between atomic and ground molecular condensates driven by two-color PA \cite{kmc00,hwd00,ggr01,jma02} have never been observed, even in  optical lattices \cite{rbm04},  although two-photon Raman coupling was recently used to observe Rabi oscillations in a degenerate Fermi gas \cite{fhm13}. Spontaneous Raman scattering by molecules presents a formidable challenge to accessing a regime in which the two-photon coupling exceeds the decoherence rate \cite{rbm04}.

Coherence seems even less accessible in one-color PA because of the short lifetime of excited molecular states. Simply increasing the laser intensity and atom-molecule Rabi frequency  does not necessarily offer a solution  because transfer of population to noncondensate atomic states becomes dominant at high laser coupling \cite{kmc00,ggr01,jma02,gas04,gas04rapid,nse06,ntj08}. (This limitation is  analogous to spontaneous Raman scattering in two-color PA.) Coherent wave-function dynamics have been observed with broadband, femtosecond PA excitation in a thermal Rb gas \cite{sme08}, but the coherence between the molecule and atom fields was not discussed.

\begin{figure}[htbp]
\includegraphics[clip=true,keepaspectratio=true,width=3.7in,trim=.35in 0in 0.in 0in]{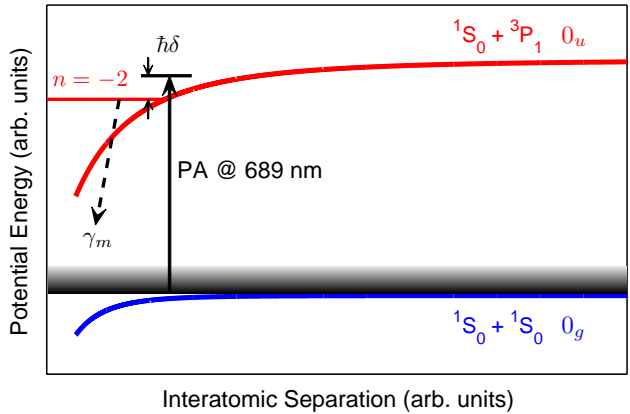}\\
\caption{(color online). One-color PA  of two ground-state $^{88}$Sr atoms in a BEC to the second least bound state ($n=-2$) of the excited molecular potential.
The shaded band indicates the continuum of scattering states important for the dynamics of Eq.\,\ref{eq:ManyBodyTheoryEqs1}.
 \label{MolecularPotential} }
\end{figure}

As pointed out in Ref.\,\cite{ntj08}, the parameter regime for coherent one-color PA using an electric-dipole allowed transition is vanishingly small, but this is not the case when exciting to a long-lived molecular state such as found on a metastable triplet potential in an alkaline-earth-metal atom.
We demonstrate this with the PA of atoms in an $^{88}$Sr condensate \cite{mmy10} to
the second least bound vibrational level on the
 $^1S_0$+$^3P_1$ molecular potential (Fig.\,\ref{MolecularPotential}) \cite{kot08,spk12}, which has  a binding energy of $h\times 24$\,MHz \cite{zbl06} with the Planck constant $h$ and transition natural linewidth of $\gamma_m/2\pi=$15\,kHz.
 Because the state is  weakly bound and the $C_3$ coefficient of the molecular potential is small, the Franck-Condon factor for the transition is large, leading to a large optical length $\ell_{\textrm{opt}}=\frac{M}{4\pi \hbar^2}\frac{|w|^2}{\hbar \gamma_m}$ \cite{eloptical}.
 $w$ is the transition matrix element between the relative-motion wave function for two atoms in the molecular and atomic-scattering channels, $\hbar$ is the reduced Planck constant, and $M$ is the atomic mass. The matrix element  is proportional to $\sqrt{I}$ for laser intensity $I$. The atom-molecule Rabi frequency is $\Omega_R=w\sqrt{n}/\hbar$, where $n$ is the condensate density.

Many-body theory describes photoassociation in a BEC   through \cite{ntj08}
\begin{eqnarray}\label{eq:ManyBodyTheoryEqs1}
i\dot{\Psi}&=&\frac{\Psi^\ast}{\hbar} \biggl(g \Psi^2 + \int \frac{\textrm{d}^3\vec{k}}{(2\pi)^3} g C_{\vec{k}}^{\textrm{dyn}}
+ w\Psi_m  \biggl), \nonumber \\
i\dot{\Psi}_m&=& \biggl(\delta^{\prime}-\delta-i\frac{\gamma_m}{2} \biggl)\Psi_m
+\frac{w}{\hbar}\Psi^2+\int \frac{\textrm{d}^3\vec{k} \,w}{(2\pi)^3\hbar}  C_{\vec{k}}^{\textrm{dyn}},\nonumber \\
&&i \dot{C}_{\vec{k}}^{\textrm{dyn}} = \frac{E_k}{\hbar} C_{\vec{k}}^{\textrm{dyn}} - i \dot{C}_{\vec{k}}^{\textrm{ad}},
\end{eqnarray}
where $\Psi$ is the atomic condensate wave function, $\Psi_m$ is the excited-state molecular condensate wave function, and $C_{\vec{k}}^{\textrm{dyn}}$ and $C_{\vec{k}}^{\textrm{ad}}$ are amplitudes for initially unpopulated scattering channels (indexed by wave vector $\vec{k}$) reflecting the dynamic and adiabatic response to the turn-on of the PA laser, respectively.
$g=4\pi \hbar^2 a/M$ describes the mean-field atom-atom interaction, where $a$ is the $s$-wave scattering length,
and $E_k=\hbar^2 k^2/M$ with the colliding-atom wave number $k$.
$\delta$ is the laser detuning from PA resonance,
and $\delta'$ is the ac Stark shift of the PA transition due to the PA laser. It arises from coupling of the excited molecular state to atomic scattering  states  besides the condensate and to  bound states of the ground molecular potential \cite{bju99}.
We have neglected elastic and inelastic collisions involving molecules, which  do not qualitatively change the results.

For a large molecular decay rate or detuning, one obtains the familiar expressions for PA loss   in the adiabatic  regime, which is incoherent and  described by a rate equation for the evolution of density $n$,  $\dot{n}=-K_{\textrm{in}} n^{2}$ for a BEC. The time-independent loss rate constant
$K_{\textrm{in}}$ is
\begin{eqnarray}\label{OFRFormulas}
K_{\textrm{in}}&=&
\frac{4\pi\hbar}{M} \frac{\ell_{\textrm{opt}}\gamma_{m}^2}{(\delta-\delta')^2+{(\gamma_{m}+\Gamma_{\textrm{stim}})^2}/{4}},
\end{eqnarray}
where $\Gamma_{\textrm{stim}} = 2k \ell_{\textrm{opt}} \gamma_{m}$ is the laser-stimulated linewidth.

On short time scales  compared  to decoherence and for $\Omega_R>\sqrt{(\delta-\delta^{\prime})^2+(\gamma_m/2)^2}$, PA is coherent and transient spectral shifts and broadenings resulting from the turn-on of the laser at $t=0$ become important.
The most striking prediction for this regime is coherent oscillations between atomic and molecular condensates, which obviously cannot be described by a rate equation.
We can obtain  Rabi frequencies as high as $\Omega_R=10^6$\,s$^{-1}$ and thus access the coherent regime for  small to moderate detuning.

Details of the formation of an $^{88}$Sr BEC are given in Ref.\,\cite{mmy10}.
We perform PA in an optical dipole trap (ODT) formed by crossed 1064 nm laser beams with waists of 66\,$\pm\, 3$\,$\mu$m, resulting in a cylindrical trapping potential with axial
frequency of 89\,$\pm\, 5$\,Hz, and radial frequency of 116\,$\pm \,10$\,Hz \cite{ycm11}.
Condensates have $N=7000-9000$ atoms, as determined via time-of-flight absorption imaging.
This is close to the critical number of condensate atoms for collapse arising from attractive interactions [scattering length $a= (-1.4\pm0.6)\,a_0$ from Ref.\,\cite{mmp08} and $a= (-2.0\pm0.3)\,a_0$ from Ref.\,\cite{skt10} with $a_0=0.053$\,nm], although uncertainty in $a$ leads to significant uncertainty in condensate properties. Using a variational calculation assuming a Gaussian density distribution \cite{dgp99}, the largest observed atom numbers constrain  $a> (-1.60\pm 0.05)\,a_0$.
For  $a= -1.6\,a_0$ and $N=8800$ atoms, and only including uncertainty from  trap geometry, the $1/\sqrt{e}$ density radius is $\sigma_0=(0.58\pm 0.04)$\,$\mu$m, and peak density is $n_\textrm{peak}=(3.0\pm 0.7)\times 10^{15}\,\mathrm{cm}^{-3}$. This  very high density
  is important for attaining large Rabi frequencies.

 The 689\,nm PA laser
has a waist of 200\,$\mu$m on the atoms.  For the highest intensity used here (2.4\,W/cm$^{2}$) the ac Stark shift of the ground state due to the PA laser is large ($\sim 10\,\mu$k),  but resulting mechanical effects can be neglected because the PA laser is applied for such a short time ($<8\,\mu$s).  The PA beam is applied while the ODT is on with a timing resolution  of 100\,ns.
 The ac Stark shift of the PA transition due to the ODT is small (30\,kHz) and constant for all PA measurements.

\begin{figure}[htbp]
\includegraphics[clip=true,keepaspectratio=true,width=3.5in,trim=0in 0in 0.in 0in]{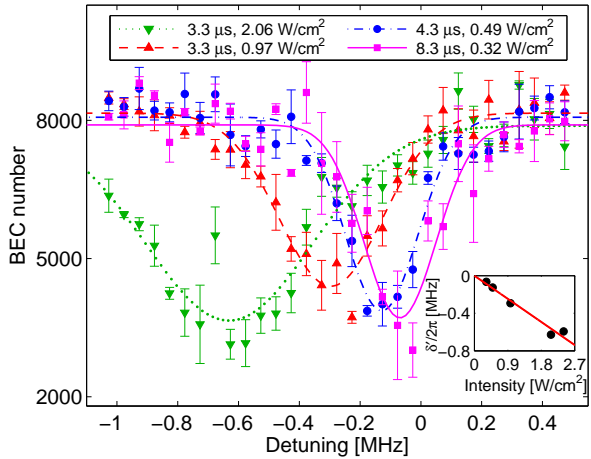}\\
\caption{(color online). ac Stark shift due to PA light for long exposure times compared to  the time scale for transient effects. Exposure times and intensities of the PA laser are given in the legend.  A fit to the line centers (inset) yields $(-275\pm 30)$\,kHz/(W/cm$^2$). Zero detuning is the line center at low intensity.
 \label{ResonanceShift} }
\end{figure}

In order to interpret the PA excitation spectra, it is important to understand the long-time and transient behavior of the ac Stark shift of the transition due to the PA laser.
Figure\,\ref{ResonanceShift} shows the shift for long excitation times, which is large compared to the linewidth for our highest intensities. Spectra are fit with  Gaussians, and our measured value of $\delta^{\prime}/(2\pi I)=(-275\pm 30)$\,kHz/(W/cm$^2$) is in agreement with the value measured in a thermal gas in Ref.\,\cite{bnb11}.
Our stated uncertainty is statistical.
There is additional systematic uncertainty arising from the transient effects on the spectra discussed below, but these effects are minimized by working at a relatively long exposure time ($>3$\,$\mu$s).

\begin{figure}[htbp]
\includegraphics[clip=true,keepaspectratio=true,width=3.5in,trim=0in 0in 0in 0in]{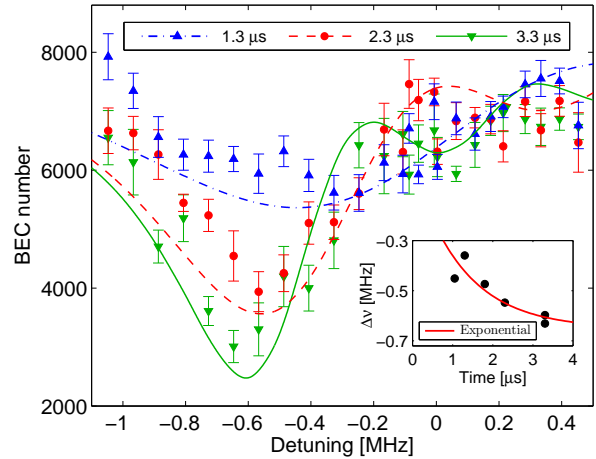}\\
\caption{(color online). Spectra of the number of atoms remaining in BEC for 2.4\,W/cm$^2$ excitation intensity and various excitation times.  Zero frequency detuning is the PA resonance position at vanishing PA laser intensity but including the ac Stark shift due to the trapping laser. Lines are predictions of
Eqs.\,\ref{eq:ManyBodyTheoryEqs1} for $\ell_{\textrm{opt}}/I$\,=\,5000$\,a_0$/(W/cm$^2$), showing the transient shift and oscillatory behavior for blue detuning.
The inset shows the  center frequency versus time, found by fitting  Gaussians to the spectra, and the solid line is an exponential fit to the trend.
 \label{BECnumVsDetuning} }
\end{figure}

Against the benchmark of the shift at long exposure times, we examine the transient behavior at shorter times. A dynamic shift at early times was predicted in Ref.\,\cite{ntj08}, and to our knowledge, we present the first experimental observation of this phenomenon.
Figure\,\ref{BECnumVsDetuning} shows a sample of high intensity PA spectra, along with an inset showing the line center as a function of exposure time. The ac Stark shift is small initially and increases on a few-microsecond time scale.

An expression for the transient behavior of the shift was derived in Ref.\,\cite{ntj08} under the sudden approximation that the amplitude in the molecular state reaches equilibrium on a much faster time scale than the observations. However, this is not valid in our experiments because the molecular-state occupation is changing during the course of the experiment.
A phenomenological fit of the trend to $\Delta\nu(t)=\Delta\nu_0\left[1-e^{-t/t_{\textrm{exp}}}\right]$ yielding $t_{\textrm{exp}}=1.24\,\mu$s is shown in Fig.\,\ref{BECnumVsDetuning} (inset).

This transient behavior has an intuitive description.
The ac Stark shift arises from the interaction between the laser electric field and the induced dipole moment reflecting coherences between molecular and continuum states. After the laser is turned on, time is required to transfer population amplitude along the internuclear axis and create coherences, and  the increasing ac Stark shift over the first few microseconds in our experiment tracks this evolution. If Eqs.\,\ref{eq:ManyBodyTheoryEqs1} are solved numerically with $C_{\vec{k}}^{\textrm{dyn}}=0$ set artificially, no transient shift is observed, but the full equations predict the shift well (Fig.\,\ref{BECnumVsDetuning}).

 The widths of the spectra in Fig.\ \ref{BECnumVsDetuning} for 1.3 and 2.3\,$\mu$s excitation closely match the expected $\sim 1/t$ widths  from energy uncertainty due to the short time of the laser pulse $t$. For 2.4\,W/cm$^2$ and the $k$ vector corresponding to the momentum width in the BEC, the stimulated width is only $\Gamma_{\mathrm{stim}}/2\pi=21$\,kHz.

Another feature  expected from an analogy with the coherent excitation of a simple two-level atomic system is oscillatory behavior, which is observed  in experiment and simulation near zero detuning in Fig.\,\ref{BECnumVsDetuning}  (around 600\,kHz blue detuning from the resonance position at long times, including ac Stark shift). In the case of coherent PA, oscillations are between atomic and molecular populations. In contrast with the atomic analogy, however, we observe distinct asymmetries and detect no oscillations at red detuning. This asymmetry is well produced in calculations of the spectra based on Eqs.\,\ref{eq:ManyBodyTheoryEqs1}, and it can be understood in the dressed atom picture. For red detuning ($\delta-\delta'<0$), the molecular state is embedded in the continuum and unstable against decay into noncondensate pairs. For blue detuning ($\delta-\delta'>0$), the molecular state is below the continuum and stable enough to contribute to oscillations. Similar behavior is expected for MFRs \cite{cgj10}.

\begin{figure}[htbp]
\includegraphics[clip=true,keepaspectratio=true,width=3.6in,trim=0.4in 0in 0in 0in]{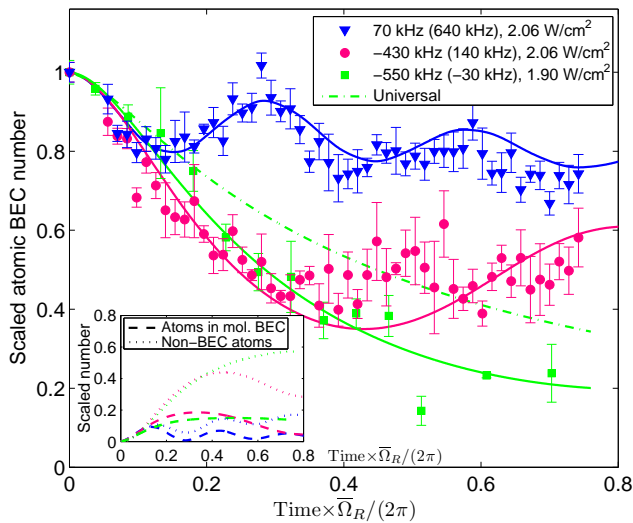}\\
\caption{(color online). Number of BEC atoms versus time. The number is scaled to the number with no PA ($\bigtriangledown$:8800, $\bigcirc$:8800, $\square$:6700) and time is scaled to the inverse Rabi frequency for average density, $\bar{n}=n_{\mathrm{peak}}/\sqrt{8}$, $2\pi/\bar{\Omega}_R=$($\bigtriangledown$:7.1, $\bigcirc$:7.1, $\square$:10.5)\,$\mu$s.  The legend states the detuning with respect to PA resonance at low intensity and (in parentheses)  the detuning with respect to the long-time resonance position including ac Stark shift, followed by the intensity. Rabi oscillations are clearest for large blue detuning from the ac-Stark-shifted resonance ($\bigtriangledown$). Solid lines are predictions of Eqs.\,\ref{eq:ManyBodyTheoryEqs1} for $\ell_{\mathrm{opt}}/I=5000\,a_0/(\mathrm{W/cm}^{2})$ and peak density at the upper limit of the measurement uncertainty. The universal prediction (dash-dotted line) is given by Eq.\,(9) in Ref.\,\cite{ntj08}. The inset shows the calculated population of atoms in the excited-state molecular BEC and ground-state atoms in noncondensate modes.
 \label{NumVsTime} }
\end{figure}

The oscillatory phenomenon is one of the most interesting aspects of our observations, as it directly manifests coherence between atomic and molecular states and has been the subject of significant theoretical attention  \cite{tth99,jma99,hwd00,icn04,kmc00,ggr01,jma02,gas04,gas04rapid,nse06,ntj08}. So, we focus on the variation of condensate population versus time for different detunings in Fig.\,\ref{NumVsTime}.
First note the initial quadratic variation with time, with loss independent of detuning, which is also a signature of the coherence of the excitation.
With increasing time, the  number at larger detuning reaches a minimum and oscillations set in.

Clear oscillations are visible for the largest blue detuning, which corresponds to the region of oscillatory behavior in Fig.\,\ref{BECnumVsDetuning}. A solution of Eqs.\,\ref{eq:ManyBodyTheoryEqs1}, which includes coupling to the continuum and associated transient effects, fits the data well. Density inhomogeneity is treated with a local density approximation. The variation in Rabi frequency resulting from the inhomogeneous density in the condensate accounts for nearly all of the reduction in contrast to the oscillations. We expect smaller contributions from
 natural decay of the molecular state \cite{zbl06,ydr13}, laser linewidth,   and atom-molecule and molecule-molecule collisions.

 The best fit in Fig.\,\ref{NumVsTime} requires a density at the upper limit of experimental uncertainty and $\ell_{\textrm{opt}}/I=(5000\pm 1000)\,a_0/(\mathrm{W/cm}^{2})$.
This is significantly less than the $\ell_{\mathrm{opt}}$ values  calculated directly from knowledge of the molecular potentials \cite{bnb11} and extracted from collisional effects of the optical Feshbach resonance in a BEC \cite{ydr13} and a thermal gas \cite{bnb11}, which is a point for further investigation.

Rabi oscillation corresponds to the return of population to the  atomic condensate, which is only possible if a coherent many-body molecular population exists. We interpret this as the presence of a molecular condensate \cite{tth99,ntj08}. The simulation  in Fig.\,\ref{NumVsTime} (inset) suggests the presence of as many as 2000 atoms in the  excited molecular electronic state remaining as an excited-state molecular condensate after the laser is turned off. The lifetime of this molecular condensate is sensitive to unknown molecule-molecule and atom-molecule inelastic collision rates as well as natural and laser-induced decay.
This represents the first demonstration of Rabi oscillations with photoassociation and
the first creation of a condensate of molecules in an excited electronic state.


Figure\,\ref{NumVsTime} (inset) also shows a significant population in atomic, noncondensate modes, which is in the form of correlated atomic pairs. This process becomes more dominant at higher laser intensity, and it ultimately limits the population of molecules \cite{kmc00,gas04}. A large atomic, noncondensate population also heralds a universal regime in which the BEC decay rate is only determined by the mass and density of
the atoms. The experimentally observed decay near resonance in Fig.\,\ref{NumVsTime}  closely matches the universal prediction of Eq.\,(9) in Ref.\,\cite{ntj08}.

The formation of a molecular condensate through coherent PA is analogous to formation using MFRs \cite{kgj06}.
Viewing the PA transition as an optical Feshbach resonance (OFR)
highlights several extreme features of the system. At detunings from resonance on the order of $\gamma_m$, the scattering length due to the OFR for the highest intensity used here is extremely large ($a\sim \ell_{\textrm{opt}}=12000$\,$a_0=640$\,nm). This is much larger than the range of the molecular potential, given by the van der Waals length $R_{\textrm{vdW}}\equiv\left(M C_6/16\hbar^2\right)^{1/4}=75$\,$a_0$, where $C_6$ is the van der Waals dispersion coefficient for the ground-state potential,
and much larger than the typical interparticle spacing $n^{-1/3}\approx 100$\,nm. This implies the system is strongly interacting and should be described as a highly correlated many-body state, rather than a gas of atoms and molecules \cite{kgj03}. For the detuning at which Rabi oscillations are observed in this study, $a\approx 500$\,$a_0$ and $na^3\ll1$.

We have presented the first experimental demonstration of coherent one-color PA, which leads to the formation of a condensate in an excited molecular state and Rabi oscillations between atomic and molecular systems.
While STIRAP \cite{bts98} and magnetoassociation \cite{kgj06} seem better suited for the practical goal of forming molecular condensates in the electronic ground state, this work shows the rich physics accessible with coherent single-color PA, such as transient effects on the excitation spectrum.

It will be very interesting to explore further the analogy between coherent PA and MFRs.  Can the oscillations in a Rabi or Ramsey configuration be used to measure the binding energy of the dressed molecular state as was done with a MFR \cite{dct02,opc09}?
 For the highest intensities used in the experiments described here, the dimensionless Feshbach resonance strength parameter \cite{cgj10} approaches unity. This suggests it may be feasible to explore  the nature of broad versus narrow optical Feshbach resonances and the relative importance of the open and closed channels on resonance \cite{gas04,ktj05,cgj10}. With OFRs, it is possible to change the resonance from narrow to broad by changing the laser intensity. There is no analogous controllability in MFRs.
 It would also be interesting to perform similar experiments in a lattice where density inhomogeneity and molecular loss due to collisions can be reduced. Photoassociative transitions involving narrower optical lines, such as optical clock transitions in strontium and ytterbium or the $^1S_0$-$^3P_1$ transition in calcium \cite{kta13}, may further expand the possibilities with coherent one-color PA.


\vspace{0 in}

\textmd{\textbf{Acknowledgements}}

This research was supported by the Welch Foundation (C-1579) and the National Science Foundation (PHY-1205946).



\begin{thebibliography}{39}
\bibitem[{\citenamefont{eloptical}(2006)\citenamefont{Zelevinsky, Boyd,
  Ludlow, Ido, Ye, Ciurylo, Naidon, and Julienne}}]{eloptical}
  \bibinfo{journal}{Published values, in $\,a_0$/(W/cm$^2$), are $\ell_{\textrm{opt}}/I = (2.2 \pm 1.0) \times 10^4$  \cite{ydr13} and $\ell_{\textrm{opt}}/I = (1.58\pm0.08) \times 10^4$ \cite{bnb11}  from experiment, and   $\ell_{\textrm{opt}}/I = 8.3 \times 10^3$ calculated from knowledge of the molecular potentials \cite{bnb11}.}
\end{thebibliography}

\end{document}